# Excited-state-induced nonlinearity and polarity reversals in the thermocurrent through a quantum dot

[1]*Parvathy Gireesan,* [2]*Giorgio Biasiol, and* [1]*Madhu Thalakulam*[*]

[1] School of Physics, Indian Institute of Science Education and Research Thiruvananthapuram, Thiruvananthapuram, Kerala 695551, India
[2] CNR-Consiglio Nazionale Delle Ricerche, Istituto Officina Dei Materiali, Area Science Park, Strada Statale 14 Km 163.5, Basovizza, Trieste, 34149, Italy

# ABSTRACT

Quantum dots are electrically tuneable thermal machines capable of both energy harvesting and cooling at the nanoscale in the cryogenic regime. At large thermal biases, their thermoelectric transport shows a complex behaviour, including additional extrema and polarity reversals in the thermocurrent. Though these asymmetries and nonlinearities directly affect rectification and power conversion, their origin has not been studied systematically. We address this through measurements and modelling of a single GaAs/AlGaAs quantum dot whose source is heated by white-noise excitation. Using Pauli master equation for sequential tunnelling through a two-level dot, retaining the full Fermi distributions of both leads, distinct tunnel rates for the two levels, and relaxation between them, we show that the excited state acts as a second transport channel which competes with the ground-state contribution. Strong excited-state coupling combined with slow relaxation produces polarity reversals in the thermocurrent; without both, the excited state yields only additional extrema and an amplitude asymmetry across the charge degeneracy point. Experimentally, as the source temperature increases, an additional extremum emerges in the thermocurrent which develops into a polarity reversal. Using the excited-state level spacing and ground-state tunnel rate obtained independently from pulsed-gate and bias spectroscopies, the model reproduces the positions of the observed extrema and their amplitudes to within a factor of about three, placing the device in the regime of slow relaxation and strong excited-state coupling. Our results establish thermocurrent line-shape as a probe of excited-state tunnelling and relaxation in quantum dots. Excited states, often treated as a minor correction, can dominate quantum-dot thermoelectric transport far from equilibrium.

---

[*]Corresponding author, madhu@iisertvm.ac.in

## I. INTRODUCTION

Quantum dots (QDs) are among the most actively studied low-dimensional systems for thermoelectric (TE) transport [1–33]. The interest is both fundamental, in quantum thermodynamics [1], and practical, since QDs can serve as on-chip, electrically tuneable thermal machines for energy harvesting [2–4] and cooling [5–7]. Their small energy scales make non-linear transport accessible at modest thermal biases, at cryogenic temperatures. In this regime, asymmetries in the TE-response allow QD devices to act as nanoscale rectifiers for charge [8,34,35] and heat [8–12,35,36] currents. This directly affects power conversion and refrigeration [13–15]. Polarity reversal and asymmetries in the TE-current also serve as a spectroscopic probe of Kondo correlations [13,16–19] and co-tunnelling [20,37,38]. Asymmetries of various physical origins have been shown, theoretically [12,14,21–23,39] and experimentally [16], to enhance the TE output. Experimentally, spin asymmetries induced by magnetic fields have improved thermal rectification through a QD [10] and the performance of a single-molecule heat engine [16]. An orbital excited state breaks electron–hole symmetry naturally at zero magnetic field, and its effect can be tuned electrically through the level spacing. Exploiting these effects, whether as a spectroscopic probe or as a design principle, requires knowing when a polarity reversal is intrinsic to the level structure of the QD.

Excited-state signatures in TE-transport were first observed as a fine structure on the Coulomb-blockade oscillations [24]. These results agreed qualitatively with the linear-response theory for QD thermopower [25], which accounts for experiments at low thermal bias [24,26–28]. At large thermal biases (comparable to the lead temperatures), the picture is far less clear. The linear-response theory [25] could not explain the suppression and polarity reversal of the thermovoltage observed at large heating currents [26]. Later experiments reported widely varying behaviour at high thermal biases: polarity reversals at specific gate voltages [15,26], additional maxima on the Coulomb-blockade oscillations [24], a fall in TE-current without polarity reversal at any gate voltages [29], and only weak deviations from linear response [40].

A theoretical description of QD thermoelectric transport beyond the linear response regime remains an open problem [1,41,42]. The Landauer–Büttiker formalism of linear TE-transport [43], extended to second order in the biases [34,35], does not produce polarity reversals or additional zeroes. Using a similar approach, the polarity reversal observed in [15] was attributed to a phenomenological fitting parameter which shifts the QD level linearly with thermal bias. In contrast, the strong non-linearity in thermoelectric transport was attributed explicitly to a competing contribution from an excited state that becomes accessible at higher temperatures, and was modelled qualitatively with a Pauli master equation [29]. Later, a two-level QD using the Anderson model and standard many-body techniques predicted polarity reversals in thermocurrent at all gate voltages between the two charge-degeneracy points [8]. These approaches [8,29] assume equal tunnel rates for the two levels and neglect relaxation to the ground state. Therefore, the conditions under which an excited state produces polarity reversals rather than additional extrema, remain unclear. The question is also difficult to settle from existing literature, because theory and experiment have rarely been confronted directly. Theoretical treatments of non-linear TE-transport [8,34,35,41,42] are generally not compared with measurements, while experimental results have been compared with linear-response theory [24,26–28], modelled qualitatively [29], or described using phenomenological fitting parameters [15].

Here, we combine modelling and experiment on the same device. We describe first-order TE-transport through a two-level QD in the sequential-tunnelling regime using the Pauli master equation [30,31,44–47]. We allow unequal tunnel rates for the two levels and relaxation of the excited state to the ground state. Since the model retains the full Fermi distribution functions of the leads instead of expanding about equilibrium, it holds for arbitrary voltage and thermal biases. We show that the excited state acts as a second transport channel, which competes with the ground-state contribution. Additional zeroes and polarity reversals occur only when the excited-state tunnel rate is faster than both the ground-state tunnel rate and the relaxation rate; they strengthen with increasing thermal bias. Otherwise, the excited state produces only additional extrema and an amplitude asymmetry across the charge-degeneracy point. We compare these theoretical findings directly with an experiment on a gate-defined GaAs/AlGaAs QD, where controllably heating the source reservoir produces an additional extremum in the TE-current that develops into a polarity reversal as the source temperature increases. With the ground-state tunnel rate and the level spacing measured independently on the same device, the model reproduces the positions of the observed extrema, and their amplitudes to within the experimental uncertainties. These results also establish the TE-current line-shape as a probe of excited-state tunnelling and relaxation in QDs.

## II. MODEL

Fig. 1 shows a schematic diagram of the system consisting of a QD with two energy levels $E_G$ and $E_E$, tunnel coupled symmetrically to left (L) and right (R) electron reservoirs, and capacitively coupled to a gate electrode of voltage $V_g$. Let $E_G$ and $E_E$ be the ground state and first excited state of the $(N+1)^{th}$ charge state with tunnel rates $\Gamma_G$ and $\Gamma_E$ respectively to the reservoirs, and $\gamma$ be the relaxation rate from $E_E$ to $E_G$. The source-drain bias $V_{SD}$ shifts the Fermi levels of the two reservoirs by $E_{SD} = \mu_L - \mu_R = -e\,V_{SD}$, where $\mu_L$ and $\mu_R$ are the electrochemical potentials of the left and right reservoirs respectively. $T_L$ and $T_R$ are the electron-temperatures of the left and right reservoirs respectively. The Fermi distribution at L(R) is given by $f_{L(R)}(\varepsilon) = [1 + \exp\left(\frac{\varepsilon - \mu_{L(R)}}{k_B T_{L(R)}}\right)]^{-1}$, where $k_B$ is the Boltzmann constant and $\varepsilon$ is the energy. We study the $N \leftrightarrow N+1$ charge transition in the regime where (i) $k_B T_L, k_B T_R < \Delta E < E_C$, (ii) tunnel barrier resistances are much higher than $h/e^2$, and (iii) $k_B T_L$ and $k_B T_R$ are larger than the intrinsic lifetime broadening of both levels [25,48,49]. $\Delta E = E_E - E_G$ is the level spacing and $E_C$ is the charging energy of the QD. In this regime, the discrete energy spectrum is well resolved and all higher-order tunnelling processes including second-order cotunneling are negligible. In addition, the quantum master equation for the open quantum system of a QD tunnel coupled to source-drain reservoirs reduces to the Pauli master equation – a set of rate equations for the time evolution of occupation probabilities of the different states possible in the system [44,45,47].

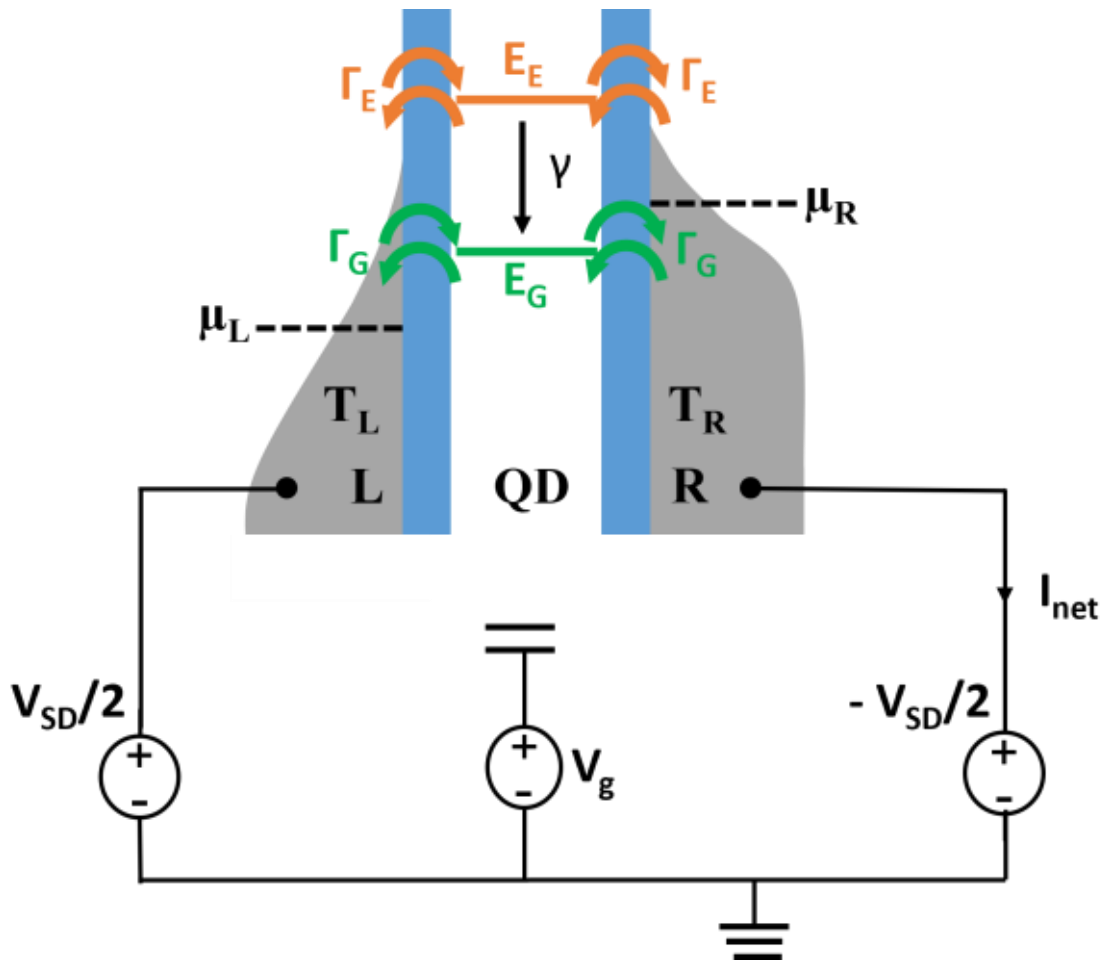

**FIG 1. Equivalent circuit of a single QD connected to source and drain electrodes.** A symmetric source-drain bias $V_{SD}$ (higher than the thermal broadening) applied galvanically to the two electron-reservoirs L/R cause a net tunnelling current $I_{net}$. The ground-state and excited-state levels are marked in green and orange, respectively. The Fermi distributions at L/R are shaded in grey, and the blue columns denote the two tunnel barriers. The tunnelling and relaxation processes are marked with their corresponding rates.

At the $N \leftrightarrow N+1$ charge transition, the three possible states of the QD are – both $E_G$ and $E_E$ being empty, $E_G$ being filled with the $(N+1)^{th}$ electron, and $E_E$ being filled with the $(N+1)^{th}$ electron. Let the occupation probabilities of these three states be $P_0$, $P_G$, and $P_E$ respectively. The transition rate for tunnelling into (out of) a QD energy level is the product of the tunnel rate to the involved lead and the probability of the relevant energy at the lead being occupied (empty). Thus, the transition rates $W_{in,G}$, $W_{in,E}$, $W_{out,G}$, and $W_{out,E}$ for tunnelling into the ground state, into the excited state, out of the ground state, and out of the excited state respectively are given by:

$$
\begin{aligned}
W_{in,G} &= \Gamma_G\, f_L(E_G) + \Gamma_G\, f_R(E_G) \\
W_{in,E} &= \Gamma_E\, f_L(E_E) + \Gamma_E f_R(E_E) \\
W_{out,G} &= \Gamma_G\,[1 - f_L(E_G)] + \Gamma_G\,[1 - f_R(E_G)] \\
W_{out,E} &= \Gamma_E\,[1 - f_L(E_E)] + \Gamma_E\,[1 - f_R(E_E)]
\end{aligned}
\qquad (1)
$$

All the transition rates are completely determined by the tunnel rates and the two Fermi-distributions. At any energy level alignment of the QD, the electron exchange between the QD and reservoirs is then described by the following rate equations:

$$
\begin{aligned}
\dot{P}_0 &= -W_{in,G}P_0 - W_{in,E}P_0 + W_{out,G}P_G + W_{out,E}P_E \\
\dot{P}_G &= +W_{in,G}P_0 - W_{out,G}P_G + \gamma P_E \\
\dot{P}_E &= +W_{in,E}P_0 - W_{out,E}P_E - \gamma P_E
\end{aligned}
\qquad (2)
$$

Solving these equations in the steady state ($\dot{P}_0 = \dot{P}_G = \dot{P}_E = 0$), along with the condition of $P_0 + P_G + P_E = 1$ which ensures single electron tunnelling, gives the corresponding steady-state occupation probabilities $P_0^{ss}, P_G^{ss}$, and $P_E^{ss}$ in terms of the transition rates.

At each tunnel barrier, the particle current from the reservoir to the QD is defined as positive. The particle currents $J_L$ and $J_R$ from the left and right reservoirs respectively to the QD are equal and opposite at steady state, and is given by

$$
J_L = +\Gamma_G\, f_L(E_G)\, P_0^{ss} + \Gamma_E\, f_L(E_E) P_0^{ss} - \Gamma_G\,[1 - f_L(E_G)] P_G^{ss} - \Gamma_E\,[1 - f_L(E_E)] P_E^{ss} \qquad (3)
$$

With L and R chosen as source and drain respectively, electron tunnelling from R to L gives the positive steady-state source-drain current in the presence of an excited state as

$$I_{net}(\varepsilon) = -e\, J_L(\varepsilon) \tag{4}$$

The energy axis $\varepsilon$ is experimentally accessed by shifting the electrochemical potential of the QD using the gate voltage $V_g$. Setting the excited state contributions $\Gamma_E\, f_L(E_E) = \Gamma_E\, f_R(E_E) = \Gamma_E\,[1 - f_L(E_E)] = \Gamma_E\,[1 - f_R(E_E)] = 0$ in Eq. (1) - (3) gives the steady-state current when transport occurs only through the ground state to be

$$I_{GS}(\varepsilon) = \frac{-e\,\Gamma_G}{2}\,[f_L(\varepsilon) - f_R(\varepsilon)] \tag{5}$$

in agreement with previous studies [2].

## III. RESULTS

### A. TE-current in the presence of a prominent excited-state

The effect of an excited state of the $(N+1)^{th}$ charge-state with $\Delta E$ = 150 μeV, at the $N \leftrightarrow N+1$ charge transition is studied in the following simulations. The total energy $\varepsilon_g$ of the QD is controlled by the gate voltage $V_g$. The position of the ground-state $E_G$ at its resonance with source-drain at $E_{SD} = 0$ is chosen as the zero of $\varepsilon_g$. A possible change to tunnel rates as the gate voltage shifts the total energy of the QD is not considered.

Fig. 2(a, b) show the Coulomb diamond plots generated with Eq. (4), for $\Gamma_G$ = 5 GHz, $\Gamma_E$ = 50 GHz, and $\gamma$ = 1 MHz. The electron temperatures at source and drain regions are kept low and equal at 10 mK in Fig. 2(a), and the Coulomb diamond shows a clear excited state line at $\Delta E$ = 150 μeV in both bias directions. Fig. 2(b) shows the corresponding Coulomb diamond plot in the presence of an additional thermal bias ($T_L$ = 500 mK, $T_R$ = 100 mK). The unequal electron-temperatures cause an unequal thermal smearing at the two slopes of the Coulomb diamond. The thermal bias causes excited-state transport to show up in QD thermocurrent at $E_{SD} = 0$ (Inset of Fig. 2(b)), while transport driven purely by source-drain bias shows only ground-state conduction at near-zero source-drain biases (Inset of Fig. 2(a)).

The evolution of excited-state features in purely thermally driven transport (at $E_{SD} = 0$, $T_L$ = 500 mK and $T_R$ = 100 mK) is shown in Fig. 2(c). The black trace shows the ground-state TE-current $I_{GS}$ (Eq. (5)), while the total TE-current $I_{net}$ due to both ground and excited states (Eq. (4)) is shown in red. The parameters used for this study are those used in Fig. 2(b), for $E_{SD} = 0$. $I_{GS}$ follows the line-shape of $f_L(\varepsilon) - f_R(\varepsilon)$ (Eq. (5)), since tunnelling is allowed through the ground-state whenever there is an imbalance in Fermi distributions across the QD [2,3,29,32]. The observed deviations of the total current in the presence of excited state from $I_{GS}$ are explained using energy level diagrams (i-vi) for the range of $\varepsilon_g$ shown in the main panel of Fig. 2(c). Ground-state (excited-state) positions are marked in black (magenta) solid lines, and the energy window where the temperature difference causes net electron tunnelling from L to R (R to L) is shaded in blue (red). $f_L(\varepsilon) - f_R(\varepsilon)$ is plotted in green along the energy axis, where the black and magenta solid circles represent the positions of $E_G$ and $E_E$. The dashed green lines mark the position $\pm(\varepsilon_0)$ of the two extrema of $f_L(\varepsilon) - f_R(\varepsilon)$. The net direction of electron tunnelling at the two levels and the relaxation from $E_E$ to $E_G$ are indicated by curved and straight purple arrows, respectively. At each energy level alignment, the resulting direction of net electron tunnelling between the two reservoirs corresponding to positive (negative) current is marked by a red (blue) solid arrow at the bottom.

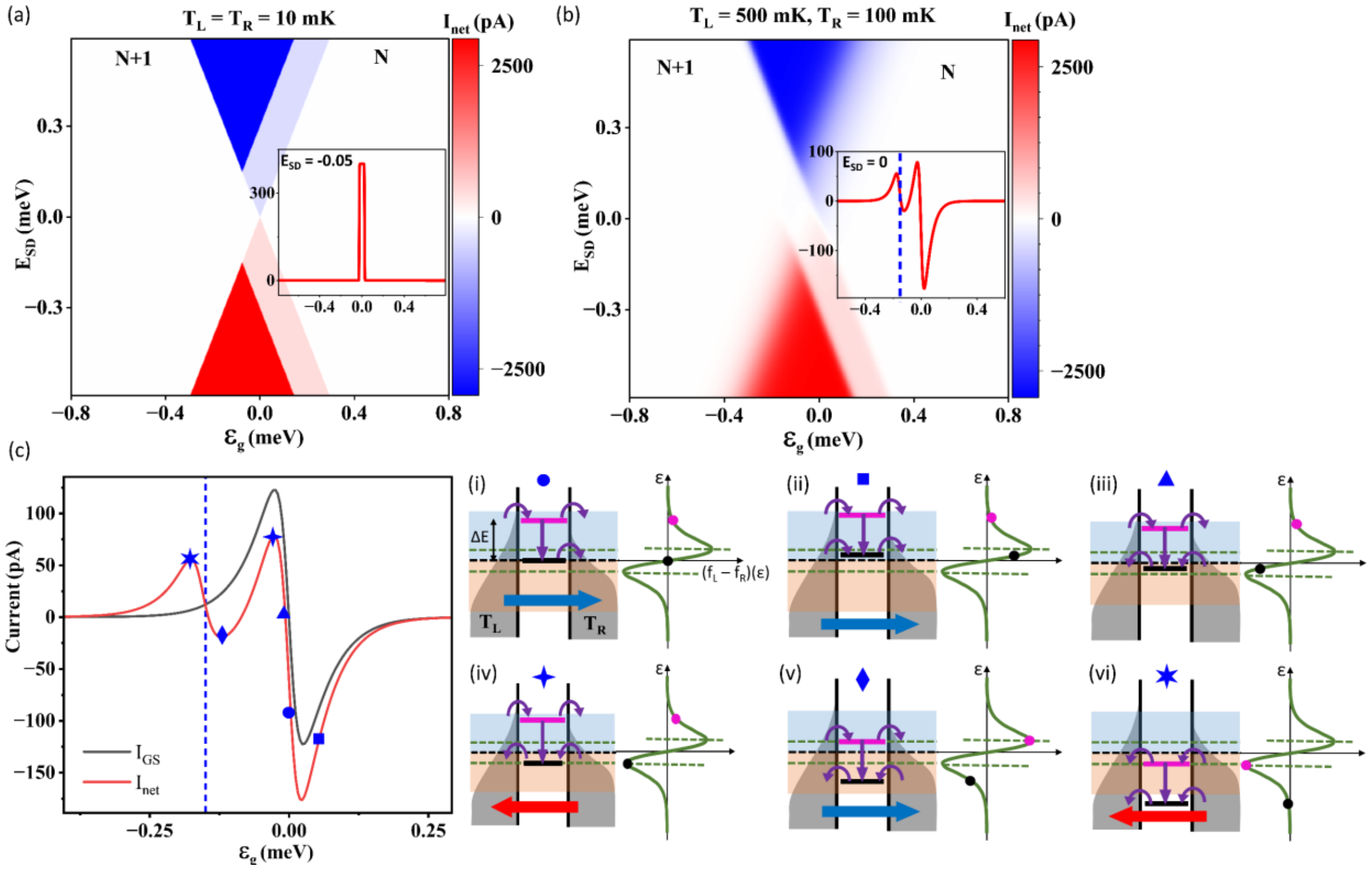

**FIG 2. Bias-driven and thermally driven transport at $\Gamma_G$ = 5 GHz, $\Gamma_E$ = 50 GHz, $\gamma$ = 1 MHz, and $\Delta E$ = 150 µeV.** Coulomb diamond plots at (a) $T_L$ = $T_R$ = 10 mK and (b) $T_L$ = 500 mK, $T_R$ = 100 mK show the excited-state lines in both bias directions. Insets of (a) and (b) show the $I_{net} - \varepsilon_g$ linecuts to the corresponding Coulomb diamonds at $E_{SD}$ values of -0.05 meV and 0 respectively. Quantities in the insets have the same units as those in their respective parent figures. (c) Ground-state ($I_{GS}$) and net ($I_{net}$) TE-currents at $T_L$ = 500 mK, $T_R$ = 100 mK, and $E_{SD}$ = 0. The energy level alignments (i-vi) of the QD corresponding to the $\varepsilon_g$ values marked in (c) explain the observed asymmetry in $I_{net}$. Blue dashed lines mark ΔE along the energy axis in (b, inset) and (c).

$I_{GS} = 0$ at $\varepsilon_g = 0$, since there is no imbalance in states between left and right at $E_G$. However, tunnelling in through the excited state $E_E$ results in a non-zero $I_{net}$ by tunnelling out either directly through $E_E$ or through $E_G$ after relaxation (Fig. 2c-(i)). The resulting $I_{net}$ is negative, since $f_L(\varepsilon) - f_R(\varepsilon) > 0$ at $E_E$. Along $\varepsilon_g > 0$, this extra contribution due to $E_E$ keeps $|I_{net}| > |I_{GS}|$ (Fig. 2c-(ii)), until $E_E$ reaches the tail of $f_L(\varepsilon) - f_R(\varepsilon)$ where $I_{net}$ falls back to $I_{GS}$. For $0 > \varepsilon_g > -\Delta E$, $E_G$ and $E_E$ fall on opposite sides of the Fermi level (Fig. 2c-(iii, iv, v)), giving opposing contributions and keeping $I_{net} < I_{GS}$.

$f_L(\varepsilon) - f_R(\varepsilon)$ exhibits a polarity reversal, and has unequal slopes on the two sides of its extrema. This difference in slopes is also manifested in $I_{GS}$ (Fig. 2(c)), which is a scaled function of $f_L(\varepsilon) - f_R(\varepsilon)$. Consequently, the opposing contributions through states with energies $E_G$ and $E_E$ can result in multiple zeroes in $I_{net}$ along $\varepsilon_g$, if the excited state contribution is strong enough. The first zero of $I_{net}$ occurs at the alignment shown in Fig. 2c-(iii), with $E_G$ at the steeper slope compared to $E_E$. As $\varepsilon_g$ is made more negative, $E_G$ aligns with $\varepsilon_0$ before $E_E$ does, since $\Delta E \gg \varepsilon_0$. $I_{net}$ thus traces out the ground-state maximum (Fig. 2c-(iv)), followed by an excited-state maximum of opposite polarity (Fig. 2c-(v)). A further decrease in $\varepsilon_g$ causes the excited-state contribution to fall more rapidly than the ground-state contribution, since $E_E$ is now on the steeper slope of $f_L(\varepsilon) - f_R(\varepsilon)$. $I_{net}$ coincides with $I_{GS}$ once $E_E$ aligns with the Fermi level, at $\varepsilon_g = -\Delta E$ (marked by blue dashed line in Fig. 2(c)). $E_E$ and $E_G$ fall on the same side of the Fermi level once $\varepsilon_g < -\Delta E$, giving $I_{net} > I_{GS}$. $I_{net}$ traces out the next excited-state maximum when $E_E$ aligns with $-\varepsilon_0$ (Fig. 2c-(vi)). $I_{net}$ falls to zero once $E_G$ reaches $f_L(\varepsilon) - f_R(\varepsilon) = 0$ and the QD enters into Coulomb blockade regime.

### B. Effect of excited-state tunnel rate and relaxation rate

Both the in and out tunnel rates through the excited state, and the relaxation rate to the ground state, $\Gamma_E$ and $\gamma$ respectively, play decisive roles in the evolution of the thermoelectric current through the QD. Fig. 3(a) and (b) show, respectively, the effect of $\Gamma_E$ and $\gamma$ on the line-shape of $I_{net}$ for the device parameters corresponding to Fig. 2(c). The overall deviations from $I_{GS}$ become more prominent when either $\Gamma_E$ is increased (Fig. 3(a)) or $\gamma$ is decreased (Fig. 3(b)), in agreement with the strengthening of excited-state transport observed in the corresponding

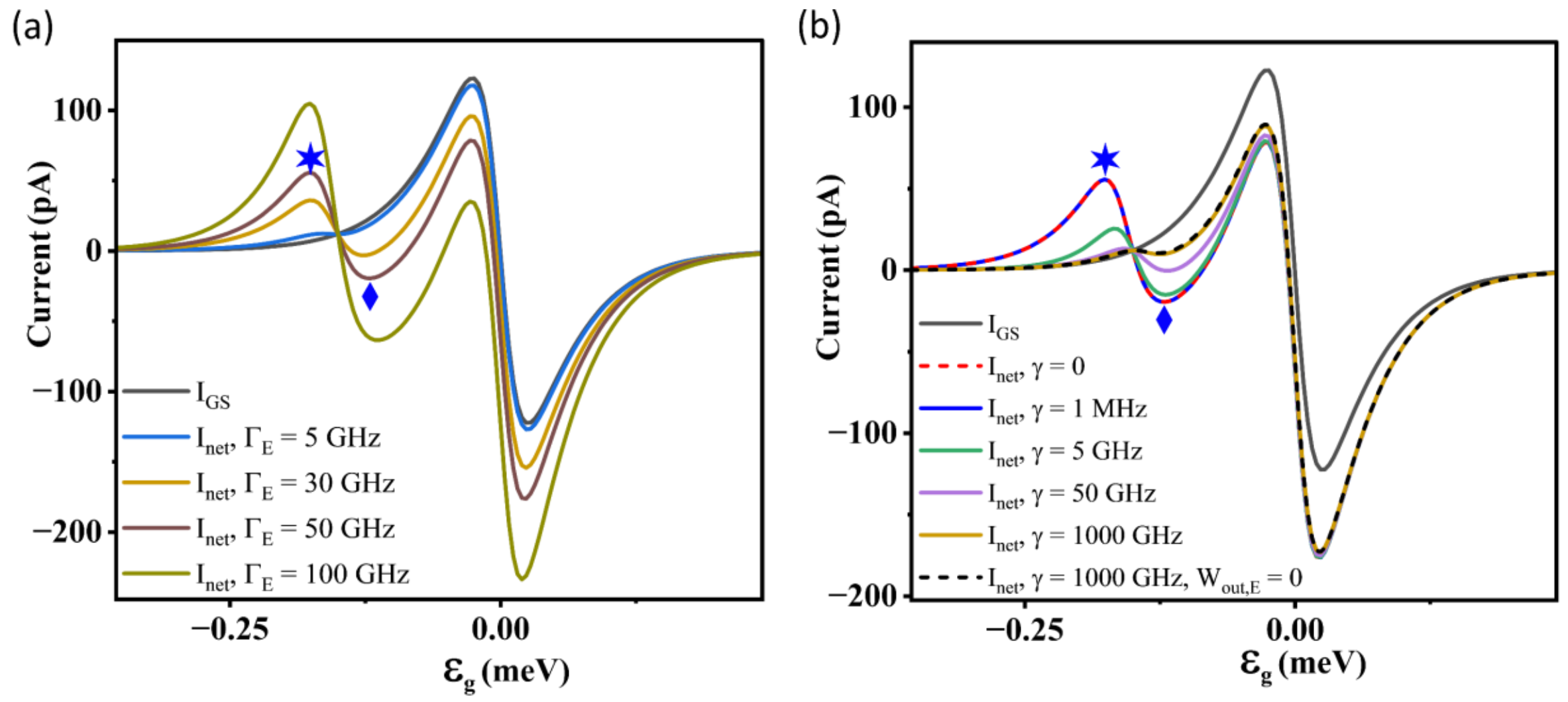

**FIG 3. Effect of $\Gamma_E$ and $\gamma$ on TE-current.** A comparison of $I_{net}$ with $I_{GS}$ at $\Gamma_G$ = 5 GHz, $\Delta E$ = 150 μeV, $T_L$ = 0.5 K, $T_R$ = 0.1 K, and $E_{SD}$ = 0 is shown for (a) varying $\Gamma_E$ at $\gamma$ = 1 MHz and (b) varying $\gamma$ at $\Gamma_E$ = 50 GHz. The excited state maxima are marked on the $\Gamma_E$ = 50 GHz and $\gamma$ = 1 MHz curves in (a) and (b). The corresponding energy level alignments at the two extrema are marked with the respective symbols in Fig. 2.

Coulomb diamond simulations in Fig. S1 of Supplemental Material [52]. However, there is a fundamental difference in the two evolutions. Decreasing $\Gamma_E$ directly reduces the excited-state contributions through both $W_{in,E}$ and $W_{out,E}$, causing $I_{net}$ to gradually evolve into $I_{GS}$ (Fig. 3(a)). On the other hand, increasing $\gamma$ at high $\Gamma_E$ relaxes the system immediately to the ground state. This drastically reduces the probability of tunnelling out from $E_E$, while maintaining the tunnelling in probability to the excited state unchanged. This is supported by the fact that the plot of $I_{net}$ at $\gamma$ = 1000 GHz obtained using Eq. (4) (solid yellow curve in Fig. 3(b)) coincides with that obtained from the model when the transition rates for tunnelling out of $E_E$ are kept zero (dashed black curve in Fig. 3(b), obtained by keeping $\Gamma_E\,[1 - f_L(E_E)] = \Gamma_E\,[1 - f_R(E_E)]$ = 0 in Eq. (1)-(3)). In the limit of $\gamma \gg \Gamma_E$, the two excited-state extrema (marked with the respective symbols in Fig. 3(b)) are thus weak, and do not cause the additional polarity reversals. However, as before, $I_{net} < I_{GS}$ for $\varepsilon_g > -\Delta E$ and $I_{net} > I_{GS}$ for $\varepsilon_g < -\Delta E$. This results in an amplitude-asymmetric $I_{net}$ which does not coincide with $I_{GS}$ at $\gamma \gg \Gamma_E$ (Fig. 3(b)).

### C. Effect of thermal bias

In Fig. 4, we inspect the role of thermal bias $\Delta T = T_L - T_R$ on the TE-current through the excited state. Fig. 4(a) and 4(b) show the evolution of the $\gamma$ = 1 MHz and $\gamma$ = 1000 GHz curves from Fig. 3(b), respectively, as $T_L$ is varied at $T_R$ = 0.1 K. The choice of $\gamma$ = 1 MHz as a low relaxation rate is justified by the fact that the excited-state contribution does not increase significantly when $\gamma$ is decreased below 1 MHz in Fig. 3(b).

Though the polarity-reversals at low $\gamma$ are found to become more prominent at higher thermal bias (Fig. 4(a)), we find that increasing the thermal bias does not induce any polarity reversals at high $\gamma$ (Fig. 4(b)). The ground-state extremum and the two excited-state extrema are marked on the TE-curves with symbols which correspond to their respective energy level diagrams from Fig. 2. Thus, $I_{net}$ shows additional zeroes and polarity reversals only at low relaxation rates (either $\gamma < \Gamma_G < \Gamma_E$ or $\Gamma_G < \gamma < \Gamma_E$), where the line-shape becomes very sensitive to the $\Gamma_E/\Gamma_G$ ratio. Fig. S2 of Supplemental Material [52] further validates this finding by showing that adjusting the other parameters will not cause polarity reversals in $I_{net}$ if either $\Gamma_E/\Gamma_G$ or $\Gamma_E/\gamma$ is too low.

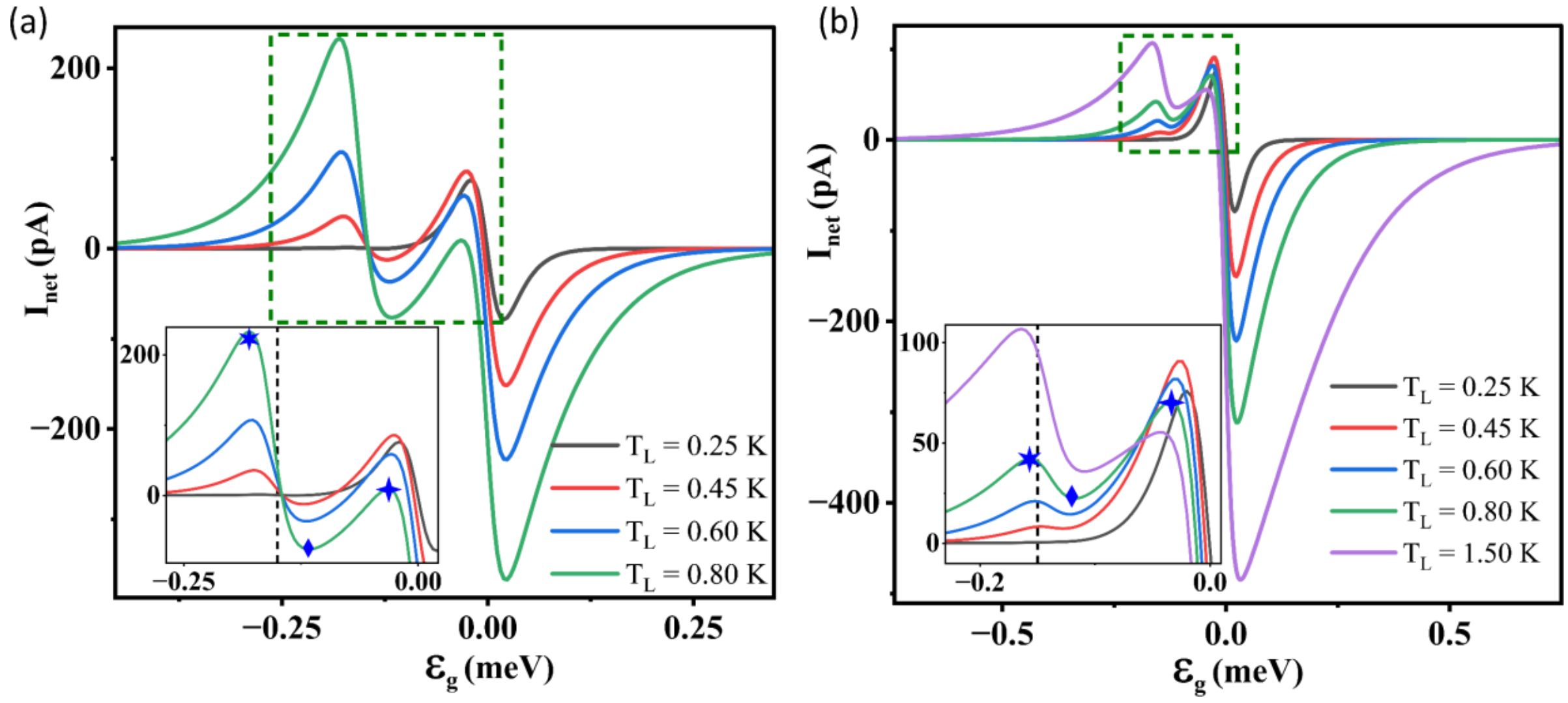

**FIG 4. Evolution of TE-current with thermal bias.** $I_{net}$ curves at $\Gamma_G$ = 5 GHz, $\Gamma_E$ = 50 GHz, $\Delta E$ = 150 μeV, and $T_R$ = 0.1 K are shown for relaxation rates of (a) 1 MHz and (b) 1000 GHz. The regimes outlined in green in (a) and (b) are shown in the respective insets. The black dashed lines in the insets mark the position of $\varepsilon_g = -\Delta E$. ✦ marks the ground-state extremum while ✶ and ♦ mark the two excited-state extrema, on the $T_L$ = 0.80 K curves in both insets. The energy level alignments at these three $\varepsilon_g$ values are marked with their corresponding symbols in Fig. 2.

## D. Experimental data

We present experimental evidence for excited-state TE-transport through a single QD defined on a doped GaAs/AlGaAs two-dimensional electron gas (2DEG). The device fabrication details are described in Section III of Supplemental Material [52], and also reported elsewhere [32]. Fig. 5(a) shows the SEM image of the depletion-gate layout of the QD along with the circuit used for measurements. The source and drain ohmic contacts are marked as S and D respectively. The device is DC voltage-biased ($V_{SD}$) in two-probe configuration, and the source-drain current $I_{SD}$ is recorded with a low-noise current pre-amplifier. All voltages are applied with respect to the drain contact, and all measurements are done in a dilution refrigerator with a base temperature ~ 10 mK. Of the six depletion gates shown in Fig. 5(a), only the four gates coloured in blue are powered to form the single QD (represented by filled yellow circle). A bias-tee attached to the gate labelled *'g'* allows simultaneous application of DC and high-frequency voltages required for conducting pulsed-gate spectroscopy. Another bias-tee at the source contact enables the application of white noise ($V_{WN}$) to induce heat and vary the electron temperature at the source. The Coulomb diamond plot corresponding to the $N \leftrightarrow N+1$ charge transition, in the absence of any applied white-noise, is shown in Fig. 5(b). We estimate a ground-state tunnel rate $\Gamma_G \approx 2.93 \pm 0.030$ GHz, from the plateaus observed in the $I_{SD} - V_{SD}$ curves near the charge degeneracy point [2,33] (see Section IV of Supplemental Material [52]). From the Coulomb diamond, we extract a lever arm of 0.0421, to calibrate $V_g$ in terms of energy $\varepsilon_g$.

Though the Coulomb diamond could reveal excited states [50], the high thermal broadening washes out the excited state features as discussed in Fig. 2(b). We independently confirm the existence of excited-states at this charge transition, using pulsed-gate voltage spectroscopy under a finite $V_{SD}$ [51]. Pulsed-gate spectroscopy at $V_{SD}$ = -17 μV (Fig. 5(c)) shows two excited-state transitions corresponding to the excited-states of N and N+1, marked with red and black dashed lines respectively. The numerical derivative $dI_{SD}/d\varepsilon_g$ of the pulsed gate spectroscopy data is used to extract the energy spacing between the ground-state (marked in red solid lines in Fig. 5(c)) and the first excited-state of the N-electron charge state to be $\Delta E$ ~ 226 μeV (see Fig. S4 of Supplemental Material [52]). $\varepsilon_g = 0$ in Fig. 5(c, d, e) corresponds to the charge degeneracy point of the transition.

The transfer characteristics of the QD close to zero source-drain bias, top panel in Fig. 5(d), shows an antisymmetric shape with a peak and a dip, a signature of TE-current [3,16,32,33]. We find the emergence of additional extrema and polarity reversals as the amplitude of white noise ($V_{WN}$) is increased, causing heating at the source side (bottom panel of Fig. 5(d)). The black dashed line in Fig. 5(d) marks the charge degeneracy point. The $I_{SD} - \varepsilon_g$ curve at $V_{WN}$ = 0 is fitted with ground-state TE-current (Eq. (5)), to obtain the electron temperatures as $T_L = 372 \pm 16$ mK and $T_R = 347 \pm 15$ mK (top panel of Fig. 5(d)).

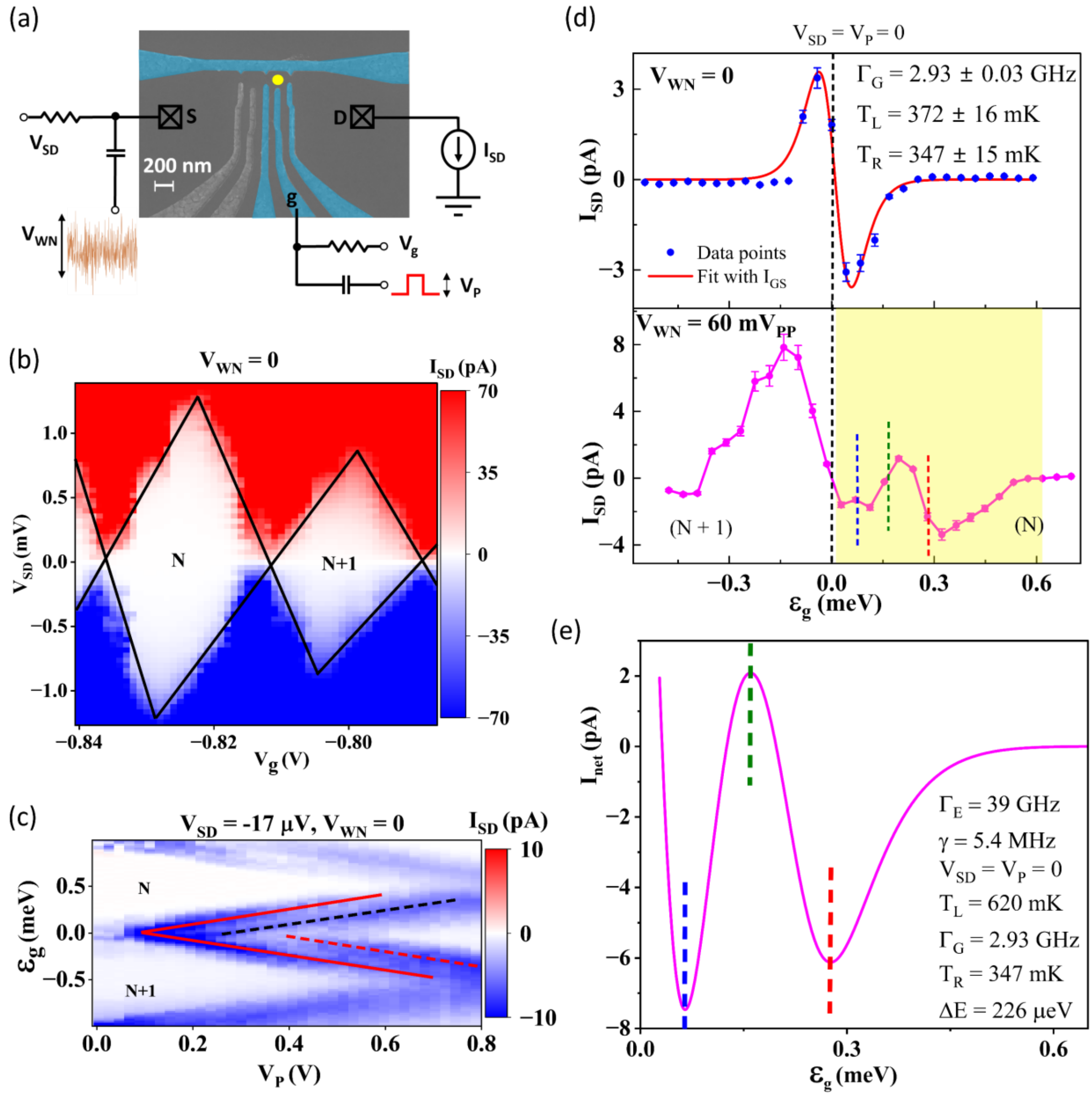

**FIG 5. Experimental evidence for excited-state TE transport through a single QD.** (a) SEM image of the depletion gate layout in the device, with the circuit used. (b) Coulomb diamond plot at $V_{WN}$ = 0 showing the $N \leftrightarrow N+1$ charge transition. (c) Bias-driven pulsed gate spectroscopy data at $V_{WN}$ = 0 showing the ground-state peak splitting (red solid lines) and excited-state transitions (dashed lines). (d) TE-current (plotted with the experimental error bars) as a function of $\varepsilon_g$, showing the evolution of excited-state induced extrema and polarity reversals at higher $V_{WN}$. The top panel of (d) shows the TE-current at $V_{WN}$ = 0 fitted to $I_{GS}$. The magenta solid line in the bottom panel of (d) is a guide to the eye. The $I_{net}$ curve simulated in (e) is seen to reproduce the $\varepsilon_g > 0$ half of the experimental TE-current at $V_{WN}$ = 60 m$V_{PP}$ (regime shaded in yellow in the bottom panel of (d)), within an order of magnitude.

Additional extrema are seen to develop on both sides of the charge degeneracy point at higher $V_{WN}$ (bottom panel of Fig. 5(d)). However, the features at $\varepsilon_g < 0$ (corresponding to the excited-state of N+1 charge state) merge with the successive resonance at higher $V_{WN}$, and cannot be interpreted accurately. Hence, we focus on the evolution of the features at $\varepsilon_g > 0$ (regime shaded in yellow in the bottom panel of Fig. 5(d)), which shows a clear separation from its neighbouring charge transition.

Since the features at $\varepsilon_g > 0$ correspond to the excited-state of the N-electron charge state, we inspect its role in the TE-current. Analogous to Eq. (3), Eq. (S5) in Supplemental Material [52] gives $I_{net}$ at the N↔N+1 charge transition of a QD when the first excited-state of N and the ground-states of N and N+1 are available for transport. We fit the $\varepsilon_g > 0$ regime of the TE-curve at $V_{WN}$ = 60 $mV_{PP}$ (regime shaded in yellow in the bottom panel of Fig. 5(d)) with Eq. (S5) of Supplemental Material. With $\Gamma_G, T_R$, and $\Delta E$ known, we use a least squares fitting approach to obtain the fit parameters to be $\Gamma_E = 39$ GHz, $\gamma = 5.4$ MHz, and $T_L = 0.62$ K (See Section VII of Supplemental Material [52] for details on the fitting procedure).

The TE-curve simulated using the fitted values of $\Gamma_E$, $\gamma$, and $T_L$ (shown in Fig. 5(e)) is seen to reproduce the corresponding experimental data at $V_{WN} = 60mV_{PP}$ (regime shaded in yellow in Fig. 5(d, bottom panel)), within an order of magnitude. The $\varepsilon_g$ values corresponding to the three extrema along the simulated curve are marked with the same colours on the experimental data, showing a very good agreement within the experimental uncertainties. We find that the data and its simulation indicate the evolution predicted by the model in Fig. 4(a) at $\Gamma_E > \Gamma_G >> \gamma$. We note that a more precise modelling of the data would require independent estimates of $T_L$ and $T_R$, which was not attempted in the experiment.

## IV. CONCLUSIONS

We have modelled thermally driven first-order transport through a two-level QD using Pauli master equation in the sequential-tunnelling regime, retaining the full Fermi distributions of the leads, distinct tunnel rates for the ground and excited states, and relaxation between them. The excited state acts as a second transport channel whose contribution opposes that of the ground state whenever the two levels lie on opposite sides of the Fermi level. Depending on $\Gamma_E / \Gamma_G$ and $\gamma$, this produces additional extrema in the TE-current, asymmetric amplitudes across the charge degeneracy point, a shift in the zero crossing of $I_{net}$ away from the charge degeneracy point, and additional zeroes accompanied by polarity reversals. The polarity reversals require $\Gamma_E \gg \Gamma_G$, and a relaxation rate which is small compared to the tunnel rate through the excited-state. They become more pronounced with increasing thermal bias, whereas fast relaxation suppresses them at any thermal bias. We have followed the TE-current along the detuning axis as the temperature of one lead is raised, rather than the more commonly studied temperature dependence at fixed detuning. This approach distinguishes the additional zeroes that emerge at higher thermal bias from an apparent shift in the charge degeneracy point.

Experimentally, in a GaAs/AlGaAs QD with a heated source electrode, we observe an excited-state extremum in the TE-current that develops into an additional zero, with polarity reversals in its vicinity, as the source temperature increases. With $\Gamma_G$ and $\Delta E$ obtained independently from the Coulomb diamond and pulsed-gate spectroscopies, the model reproduces the positions of the extrema, and their amplitudes to within a factor of about three, for $\Gamma_E \approx 13\, \Gamma_G$ and a relaxation rate at least three orders of magnitude below both tunnel rates. This places the device in the regime identified above. Our results establish the TE-current lineshape as a probe of excited-state tunnelling and relaxation in quantum dots.

**ACKNOWLEDGEMENTS**

M.T. acknowledges funding support from the National Quantum Mission, an initiative of the Department of Science and Technology, Government of India, under the Grant No. DST/QTC/NQM/QC/2024.

**REFERENCES**

[1] S. Campbell et al., Roadmap on quantum thermodynamics, Quantum Sci. Technol. **11**, 012501 (2026).

[2] M. Esposito, K. Lindenberg, and C. Van Den Broeck, Thermoelectric efficiency at maximum power in a quantum dot, Europhys. Lett. **85**, 60010 (2009).

[3] M. Josefsson, A. Svilans, A. M. Burke, E. A. Hoffmann, S. Fahlvik, C. Thelander, M. Leijnse, and H. Linke, A quantum-dot heat engine operating close to the thermodynamic efficiency limits, Nature Nanotech **13**, 920 (2018).

[4] M. Josefsson, A. Svilans, H. Linke, and M. Leijnse, Optimal power and efficiency of single quantum dot heat engines: Theory and experiment, Phys. Rev. B **99**, 235432 (2019).

[5] H. L. Edwards, Q. Niu, and A. L. De Lozanne, A quantum-dot refrigerator, Applied Physics Letters **63**, 1815 (1993).

[6] J. R. Prance, C. G. Smith, J. P. Griffiths, S. J. Chorley, D. Anderson, G. A. C. Jones, I. Farrer, and D. A. Ritchie, Electronic Refrigeration of a Two-Dimensional Electron Gas, Phys. Rev. Lett. **102**, 146602 (2009).

[7] Y. Zhang, G. Lin, and J. Chen, Three-terminal quantum-dot refrigerators, Phys. Rev. E **91**, 052118 (2015).

[8] M. A. Sierra and D. Sánchez, Strongly nonlinear thermovoltage and heat dissipation in interacting quantum dots, Phys. Rev. B **90**, 115313 (2014).

[9] M. A. Sierra and D. Sánchez, Nonlinear Heat Conduction in Coulomb-blockaded Quantum Dots, Materials Today: Proceedings **2**, 483 (2015).

[10] R. Scheibner, M. König, D. Reuter, A. D. Wieck, C. Gould, H. Buhmann, and L. W. Molenkamp, Quantum dot as thermal rectifier, New J. Phys. **10**, 083016 (2008).

[11] D. M.-T. Kuo and Y. Chang, Thermoelectric and thermal rectification properties of quantum dot junctions, Phys. Rev. B **81**, 205321 (2010).

[12] M. Wierzbicki and R. Świrkowicz, Electric and thermoelectric phenomena in a multilevel quantum dot attached to ferromagnetic electrodes, Phys. Rev. B **82**, 165334 (2010).

[13] D. Sánchez and R. López, Nonlinear phenomena in quantum thermoelectrics and heat, Comptes Rendus. Physique **17**, 1060 (2016).

[14] V. Talbo, J. Saint-Martin, S. Retailleau, and P. Dollfus, Non-linear effects and thermoelectric efficiency of quantum dot-based single-electron transistors, Sci Rep **7**, 14783 (2017).

[15] S. F. Svensson, E. A. Hoffmann, N. Nakpathomkun, P. M. Wu, H. Q. Xu, H. A. Nilsson, D. Sánchez, V. Kashcheyevs, and H. Linke, Nonlinear thermovoltage and thermocurrent in quantum dots, New J. Phys. **15**, 105011 (2013).

[16] S. Volosheniuk, R. Conte, E. Pyurbeeva, T. Baum, M. Vilas-Varela, S. Fernández, D. Peña, H. S. J. Van Der Zant, and P. Gehring, A Single-Molecule Quantum Heat Engine, Nano Lett. **26**, 984 (2026).

[17] A. Svilans, M. Josefsson, A. M. Burke, S. Fahlvik, C. Thelander, H. Linke, and M. Leijnse, Thermoelectric Characterization of the Kondo Resonance in Nanowire Quantum Dots, Phys. Rev. Lett. **121**, 206801 (2018).

[18] R. Scheibner, H. Buhmann, D. Reuter, M. N. Kiselev, and L. W. Molenkamp, Thermopower of a Kondo Spin-Correlated Quantum Dot, Phys. Rev. Lett. **95**, 176602 (2005).

[19] B. Dutta, D. Majidi, A. García Corral, P. A. Erdman, S. Florens, T. A. Costi, H. Courtois, and C. B. Winkelmann, Direct Probe of the Seebeck Coefficient in a Kondo-Correlated Single-Quantum-Dot Transistor, Nano Lett. **19**, 506 (2019).

[20] R. Scheibner, E. G. Novik, T. Borzenko, M. König, D. Reuter, A. D. Wieck, H. Buhmann, and L. W. Molenkamp, Sequential and cotunneling behavior in the temperature-dependent thermopower of few-electron quantum dots, Phys. Rev. B **75**, 041301 (2007).

[21] T. A. Costi and V. Zlatić, Thermoelectric transport through strongly correlated quantum dots, Phys. Rev. B **81**, 235127 (2010).

[22] B. Dong and X. L. Lei, Effect of the Kondo correlation on the thermopower in a quantum dot, J. Phys.: Condens. Matter **14**, 11747 (2002).

[23] E. Pyurbeeva and R. Kosloff, Quantum Dot Thermal Machines—A Guide to Engineering, Entropy **28**, 2 (2025).

[24] A. S. Dzurak, C. G. Smith, C. H. W. Barnes, M. Pepper, L. Martín-Moreno, C. T. Liang, D. A. Ritchie, and G. A. C. Jones, Thermoelectric signature of the excitation spectrum of a quantum dot, Phys. Rev. B **55**, R10197 (1997).

[25] C. W. J. Beenakker and A. A. M. Staring, Theory of the thermopower of a quantum dot, Phys. Rev. B **46**, 9667 (1992).

[26] A. A. M. Staring, L. W. Molenkamp, B. W. Alphenaar, H. V. Houten, O. J. A. Buyk, M. A. A. Mabesoone, C. W. J. Beenakker, and C. T. Foxon, Coulomb-Blockade Oscillations in the Thermopower of a Quantum Dot, Europhys. Lett. **22**, 57 (1993).

[27] A. S. Dzurak, C. G. Smith, M. Pepper, D. A. Ritchie, J. E. F. Frost, G. A. C. Jones, and D. G. Hasko, Observation of Coulomb blockade oscillations in the thermopower of a quantum dot, Solid State Communications **87**, 1145 (1993).

[28] L. Molenkamp, A. A. M. Staring, B. W. Alphenaar, H. V. Houten, and C. W. J. Beenakker, Sawtooth-like thermopower oscillations of a quantum dot in the Coulomb blockade regime, Semicond. Sci. Technol. **9**, 903 (1994).

[29] A. Svilans, A. M. Burke, S. F. Svensson, M. Leijnse, and H. Linke, Nonlinear thermoelectric response due to energy-dependent transport properties of a quantum dot, Physica E: Low-Dimensional Systems and Nanostructures **82**, 34 (2016).

[30] P. Senapati and P. Parida, Thermoelectric performance of quantum dots embedded in an Aharonov-Bohm ring: a Pauli master equation approach, Sci Rep **15**, 13232 (2025).

[31] M. Bagheri Tagani and H. Rahimpour Soleimani, Thermoelectric effects through weakly coupled double quantum dots, Physica B: Condensed Matter **407**, 765 (2012).

[32] P. Gireesan, U. Soni, H. Sundaresan, G. Biasiol, and M. Thalakulam, Tunnel-rate controlled local heat distribution in mesoscopic circuits, Quantum Sci. Technol. **11**, 035046 (2026).

[33] S. Dorsch, A. Svilans, M. Josefsson, B. Goldozian, M. Kumar, C. Thelander, A. Wacker, and A. Burke, Heat Driven Transport in Serial Double Quantum Dot Devices, Nano Letters **21**, 988 (2021).

[34] D. Sánchez and R. López, Scattering Theory of Nonlinear Thermoelectric Transport, Phys. Rev. Lett. **110**, 026804 (2013).

[35] J. Meair and P. Jacquod, Scattering theory of nonlinear thermoelectricity in quantum coherent conductors, J. Phys.: Condens. Matter **25**, 082201 (2013).

[36] R. López and D. Sánchez, Nonlinear heat transport in mesoscopic conductors: Rectification, Peltier effect, and Wiedemann-Franz law, Phys. Rev. B **88**, 045129 (2013).

[37] K. A. Matveev and A. V. Andreev, Thermopower of a single-electron transistor in the regime of strong inelastic cotunneling, Phys. Rev. B **66**, 045301 (2002).

[38] M. Turek and K. A. Matveev, Cotunneling thermopower of single electron transistors, Phys. Rev. B **65**, 115332 (2002).

[39] M. Leijnse, M. R. Wegewijs, and K. Flensberg, Nonlinear thermoelectric properties of molecular junctions with vibrational coupling, Phys. Rev. B **82**, 045412 (2010).

[40] P. Reddy, S.-Y. Jang, R. A. Segalman, and A. Majumdar, Thermoelectricity in Molecular Junctions, Science **315**, 1568 (2007).

[41] S. H. Dong, K. X. Jia, X. Y. Liu, H. Geng, L. Sheng, and D. Y. Xing, Nonlinear Landauer formula for thermal transport of the electrons, EPL **145**, 56001 (2024).

[42] G. Diniz, S. Quintino, and V. V. França, Transport in Single Quantum Dots: A Review from Linear Response to Nonlinear Regimes, Braz J Phys **56**, 27 (2026).

[43] P. N. Butcher, Thermal and electrical transport formalism for electronic microstructures with many terminals, J. Phys.: Condens. Matter **2**, 4869 (1990).

[44] F. Campaioli, J. H. Cole, and H. Hapuarachchi, Quantum Master Equations: Tips and Tricks for Quantum Optics, Quantum Computing, and Beyond, PRX Quantum **5**, 020202 (2024).

[45] R. A. Bush, E. D. Ochoa, and J. K. Perron, Transport through quantum dots: An introduction via master equation simulations, American Journal of Physics **89**, 300 (2021).

[46] P. Senapati and P. Parida, Charge and spin thermoelectric transport in benzene-based molecular nano-junctions: a quantum many-body study, Nanoscale **16**, 2574 (2024).

[47] P.-Y. Yang and W.-M. Zhang, Master Equation Approach to Transient Quantum Transport in Nanostructures, Front. Phys. **12**, 127204 (2017).

[48] C. W. J. Beenakker, Theory of Coulomb-blockade oscillations in the conductance of a quantum dot, Phys. Rev. B **44**, 1646 (1991).

[49] E. Bonet, M. M. Deshmukh, and D. C. Ralph, Solving rate equations for electron tunneling via discrete quantum states, Phys. Rev. B **65**, 045317 (2002).

[50] L. P. Kouwenhoven, D. G. Austing, and S. Tarucha, Few-electron quantum dots, Rep. Prog. Phys. **64**, 701 (2001).

[51] J. M. Elzerman, R. Hanson, L. H. Willems Van Beveren, L. M. K. Vandersypen, and L. P. Kouwenhoven, Excited-state spectroscopy on a nearly closed quantum dot via charge detection, Applied Physics Letters **84**, 4617 (2004).

[52] See Supplemental Material

# Supplemental Material: Excited-state-induced nonlinearity and polarity reversals in the thermocurrent through a quantum dot

[1]*Parvathy Gireesan,* [2]*Giorgio Biasiol, and* [1]*Madhu Thalakulam*[*]

[1] School of Physics, Indian Institute of Science Education and Research Thiruvananthapuram, Thiruvananthapuram, Kerala 695551, India
[2] CNR-Consiglio Nazionale Delle Ricerche, Istituto Officina Dei Materiali, Area Science Park, Strada Statale 14 Km 163.5, Basovizza, Trieste, 34149, Italy

## I. Coulomb diamond simulations at varying tunnel rates and γ

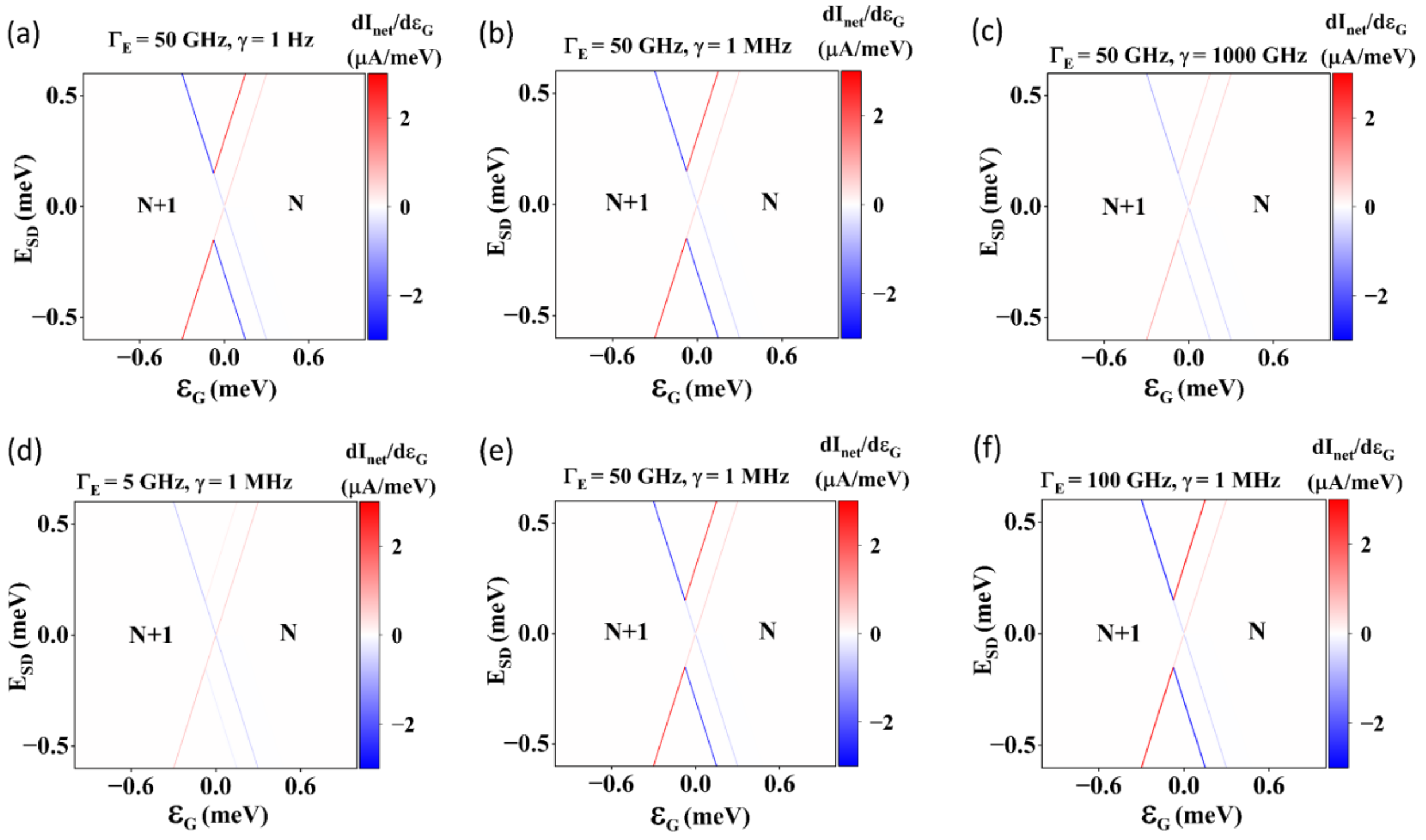

**FIG S1. Coulomb diamond simulations in $dI_{net}/d\varepsilon_g$at $\Gamma_G$ = 5 GHz and $T_L = T_R$ = 10 mK.** $\gamma$ is varied at $\Gamma_E$ = 50 GHz in (a, b, c), while $\Gamma_E$ is varied at $\gamma$ = 1 MHz in (d, e, f). The simulations are done for the $N \leftrightarrow N+1$ charge transition, with an excited state of N+1 at $\Delta E$ = 150 μeV.

Fig. S1 shows the response of the ground-state and excited-state transitions to $\Gamma_E$ and $\gamma$, when transport is purely bias-driven at $\Gamma_G$ = 5 GHz and $T_L = T_R$ = 10 mK. The excited-state resonance lines (at both source and drain) become more prominent when excited-state transport is

[*]Corresponding author, madhu@iisertvm.ac.in

increased – by either decreasing $\gamma$ at constant $\Gamma_E$ (Fig. S1(a-c)) or increasing $\Gamma_E$ at constant $\gamma$ (Fig. S1(d-f)).

### II. Effect of thermal bias at low ΓE/ΓG and low ΓE/γ

Fig. S2 demonstrates the validity of the criteria that both $\Gamma_E/\Gamma_G$ and $\Gamma_E/\gamma$ ratios should be high for $I_{net}$ to show polarity reversals, irrespective of the thermal bias applied. Increasing thermal bias does not result in polarity reversals even at $\gamma = 0$, if $\Gamma_E = \Gamma_G$ as in Fig. S2(a). Nor does it happen for the case of $\Gamma_E > \Gamma_G$, if $\gamma \gg \Gamma_E$ as in Fig. S2(b).

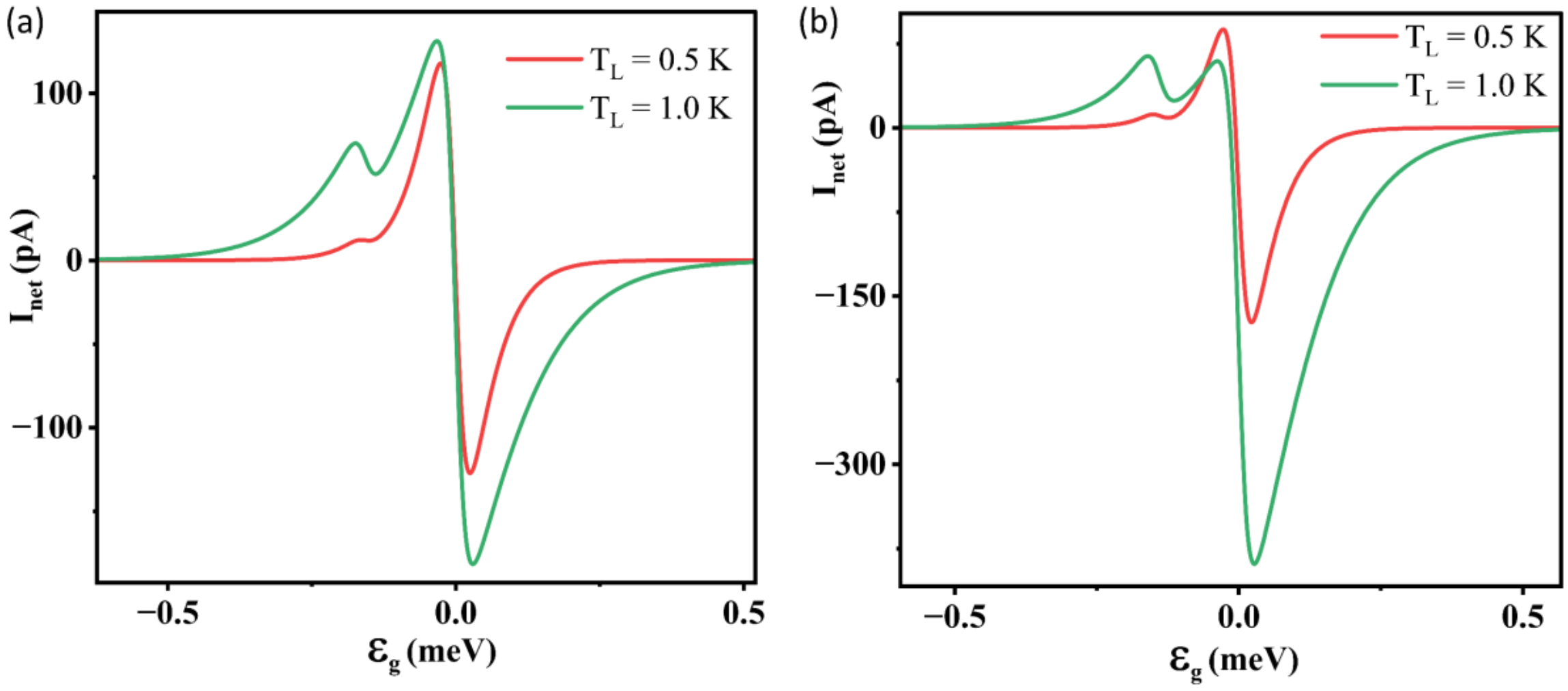

**FIG S2. TE-curves at low $\Gamma_E/\Gamma_G$ and low $\Gamma_E/\gamma$.** $I_{net}$ curves at $\Delta E$ = 150 μeV and $T_R$ = 0.1 K for (a) $\Gamma_G = \Gamma_E$ = 5 GHz, $\gamma$ = 0, and (b) $\Gamma_G$ = 5 GHz, $\Gamma_E$ = 50 GHz, $\gamma$ = 500 GHz.

### III. Device fabrication

The device is fabricated on a doped GaAs/AlGaAs heterostructure, hosting a 2DEG, ~ 90 nm below the wafer surface with a carrier density and mobility of ~ 1.75 x $10^{11}$ $cm^{-2}$ and ~ 1.5 x $10^{6}$ $cm^2/Vs$ respectively at 4 K. The wafer is wet-etched with citric acid/$H_2O_2$ to define the device mesa. Indium hot joints made to the wafer are annealed at 400 $^0$C in forming-gas atmosphere to achieve ohmic contacts to the 2DEG. The QD is fabricated by split-gate technique using photolithography and electron beam lithography, followed by Cr/Au metallization. The sample is mounted on the mixing chamber plate of a cryogen-free dilution refrigerator, and all experiments are done at the base temperature of ~ 10 mK.

## IV. Experimental data - Extracting $\Gamma_G$ from Coulomb diamond

For first-order sequential tunnelling through a single-level QD, the bias-driven current saturates for $|V_{SD}| > k_B T/e$. These saturation currents ($I_{+/-}$) correspond to the first step in the $I_{SD} - V_{SD}$ curve of the QD after Coulomb blockade is lifted in each biasing direction, and are given by

$$I_+ = 2e \frac{\Gamma_L \Gamma_R}{2\Gamma_R + \Gamma_L} \quad \text{(S1)}$$

$$I_- = -2e \frac{\Gamma_L \Gamma_R}{\Gamma_R + 2\Gamma_L} \quad \text{(S2)}$$

where $\Gamma_L$ ($\Gamma_R$) is the tunnel coupling of the energy level to the left (right) electron reservoir [2, 33]. A linecut to the Coulomb diamond in Fig. 5(b) at $V_g$ = -0.8086 V gives the saturation currents as $I_+ = 400 \pm 37$ pA and $I_- = -257 \pm 3$ pA (see Fig. S3). The effective tunnel rate $\Gamma_{eff} = \frac{\Gamma_L \Gamma_R}{\Gamma_L + \Gamma_R}$ across the QD is thus obtained to be $1.465 \pm 0.015$ GHz. Since the model assumes symmetric tunnel rates, we use $\Gamma_G = 2\,\Gamma_{eff} = 2.93 \pm 0.030$ GHz in further calculations.

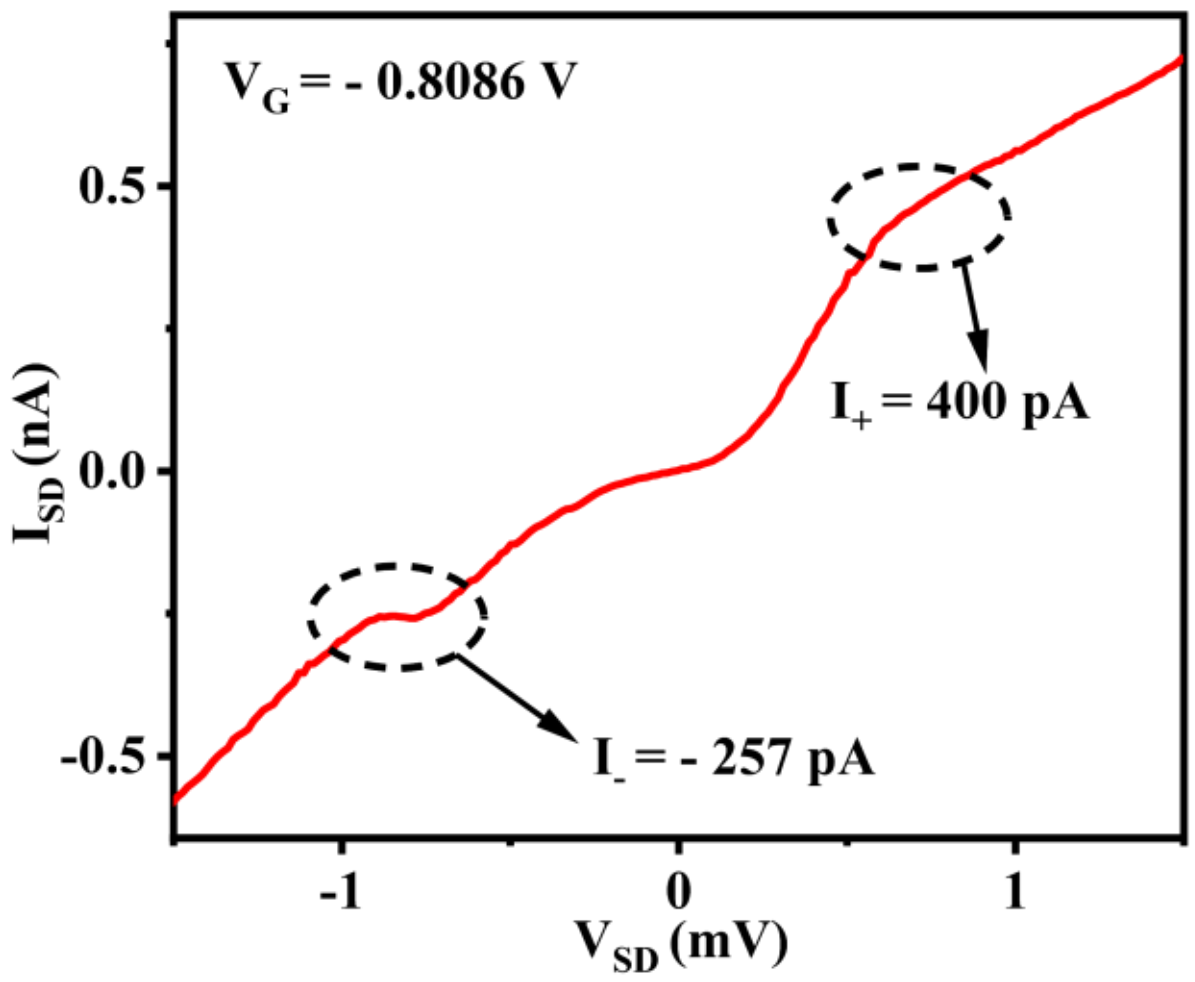

**FIG S3. Extracting $\Gamma_G$ from the $I_{SD} - V_{SD}$ curve.** Linecut to the Coulomb diamond in Fig. 5(b) at $V_g$= -0.8086 V is shown. The first plateau along the two biasing directions gives the saturation currents $I_{+/-}$.

## V. Experimental data – Extracting energy scales from pulsed gate spectroscopy data in bias-driven transport

Fig. S4 shows the pulsed gate spectroscopy data of Fig. 5(c), when plotted in the derivative $dI_{SD}/d\varepsilon_g$. The green solid lines track the peak centres of the two ground-state branches from the $dI_{SD}/d\varepsilon_g$ = 0 trace. With the excited-state line of N-electron charge state tracked similarly by the green dashed line, the spacing marked by the black solid line gives the excited-state energy as $\Delta E$ = 226 μeV.

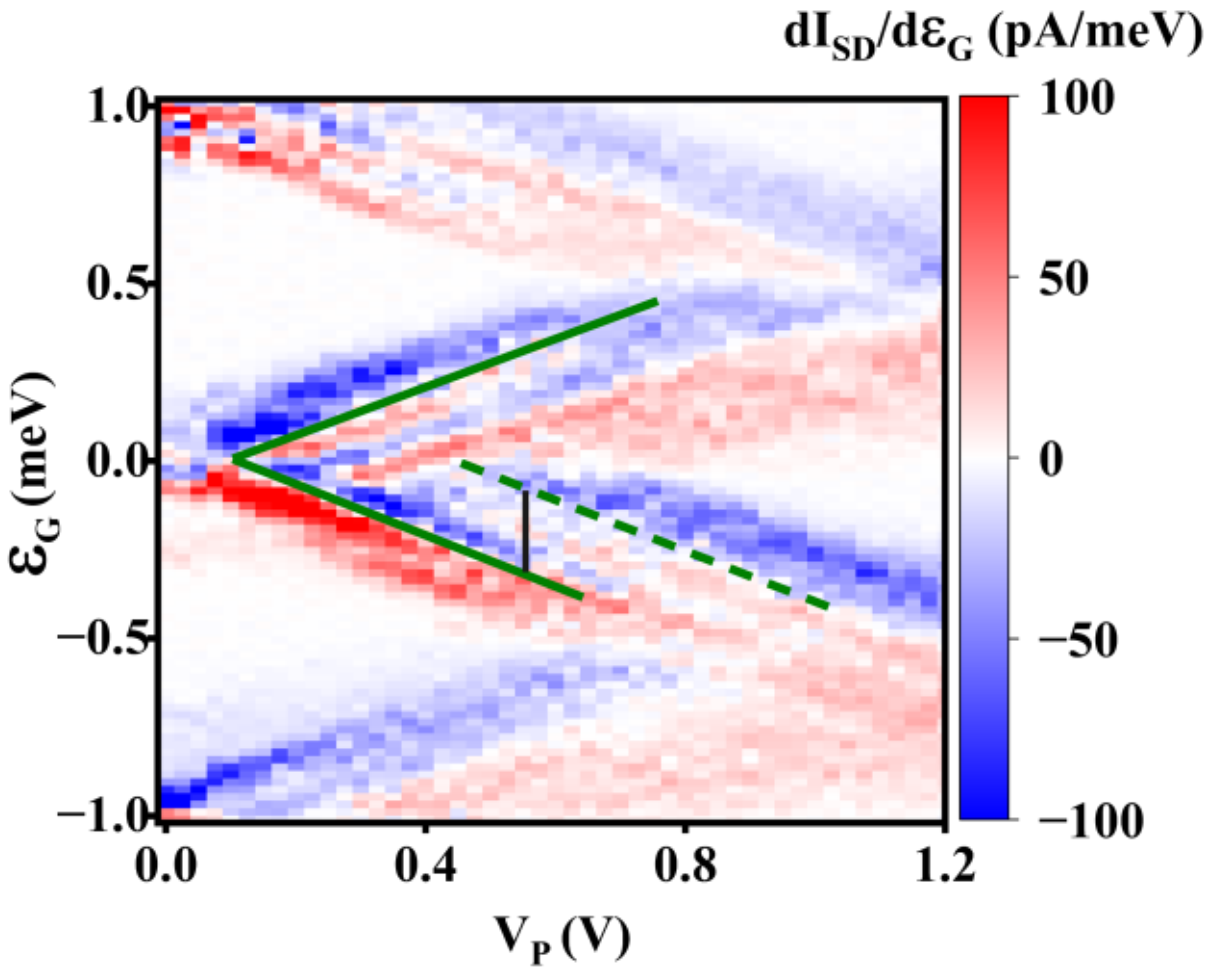

**FIG S4. Extracting energy scales from bias-driven pulsed gate spectroscopy data.** Pulsed gate spectroscopy data at $V_{SD}$ = -17 μV and $V_{WN}$ = 0, plotted in the derivative of $I_{SD}$ with respect to $\varepsilon_g$. The green solid (green dashed) lines track the peak centres of the ground-state line (excited-state line of N-electron state). The black solid line gives $\Delta E$ = 226 μeV.

## VI. Model for QD transport at the N↔N+1 charge transition when the first excited-state of N is involved

We model the transport through a QD at the $N \leftrightarrow N+1$ charge transition when the first excited-state of N and the ground-states of N and N+1 are available, using the Pauli master equation approach. $T_{L/R}$, $\mu_{L/R}$, $V_{SD}$, $E_{SD}$, $V_g$, $\varepsilon_g$, $I_{net}$, $f_{L/R}$, and $E_C$ are as defined in the main text.

Let $P_0, P_G$, and $P_E$ be the probabilities of the QD being in the (N+1)-electron ground-state, N-electron ground-state, and N-electron excited-state respectively. In this case, the (N+1)th electron tunnelling in to the QD must result in the ground-state. But, the (N+1)th electron tunnelling out of the QD at energies $E_G$ and $E_E$ result in the N-electron ground-state and N-electron excited-state respectively. $E_G$ and $E_E$ are symmetrically tunnel coupled to the reservoirs by tunnel rates at $\Gamma_G$ and $\Gamma_E$ respectively. $\Delta E$ is defined as $E_E - E_G$. The set of possible transitions thus restricts the rate equations to:

$$\begin{aligned}
\dot{P}_0 &= -W_{out,G}P_0 - W_{out,E}P_0 + W_{in,G}P_G + W_{in,E}P_E \\
\dot{P}_G &= +W_{out,G}P_0 - W_{in,G}P_G + \gamma P_E \\
\dot{P}_E &= +W_{out,E}P_0 - W_{in,E}P_E - \gamma P_E
\end{aligned} \tag{S3}$$

where $W_{in,G}$, $W_{in,E}$, $W_{out,G}$, and $W_{out,E}$ are the transition rates from N-electron ground-state to (N+1)-electron ground-state, from N-electron excited-state to (N+1)-electron ground-state, from (N+1)-electron ground-state to N-electron ground-state, and from (N+1)-electron ground-state to N-electron excited-state respectively, and $\gamma$ is the relaxation rate from N-electron excited-state to N-electron ground-state. The four transition rates are given by

$$\begin{aligned}
W_{in,G} &= \Gamma_G f_L(E_G) + \Gamma_G f_R(E_G) \\
W_{in,E} &= \Gamma_E f_L(E_E) + \Gamma_E f_R(E_E) \\
W_{out,G} &= \Gamma_G [1 - f_L(E_G)] + \Gamma_G [1 - f_R(E_G)] \\
W_{out,E} &= \Gamma_E [1 - f_L(E_E)] + \Gamma_E [1 - f_R(E_E)]
\end{aligned} \tag{S4}$$

Solving the rate equations in the steady state ($\dot{P}_0 = \dot{P}_G = \dot{P}_E = 0$), along with the normalisation condition of $P_0 + P_G + P_E = 1$ gives the steady-state occupation probabilities $P_0^{ss}$, $P_G^{ss}$, and $P_E^{ss}$ in terms of the transition rates. The resulting particle current $J_L(J_R)$ from the left (right) reservoir to the QD are equal and opposite at steady state, and is obtained to be

$$J_L = +\Gamma_G f_L(E_G)\, P_G^{ss} + \Gamma_E f_L(E_E) P_E^{ss} - \Gamma_G [1 - f_L(E_G)] P_0^{ss} - \Gamma_E [1 - f_L(E_E)] P_0^{ss} \tag{S5}$$

Choosing L and R as source and drain respectively, electron tunnelling from R to L gives positive source-drain current $I_{net}$. Hence,

$$I_{net}(\varepsilon) = -e\, J_L(\varepsilon) \tag{S6}$$

where $J_L(\varepsilon)$ is given by Eq. (S5).

## VII. Details of the least-squares fitting procedure used

The fitting is done using a Python script, which solves the steady-state rate equations given by Eq. (S3) to calculate $I_{net}$ (given by Eq. (S5) and Eq. (S6)), which is then used within a least-squares fitting algorithm. Since Sections A through C in the main text imply that observing polarity reversals in thermocurrent requires the ratios $\Gamma_E/\Gamma_G$ and $\Gamma_E/\gamma$ to be high, we allow $\Gamma_E$ and $\gamma$ to vary over the ranges 3 – 100 GHz and 0.1 MHz – 30 GHz respectively, during fitting. Based on our results reported in [32] for a similar experimental setup, we constrain $T_L$ to vary over the range 0.6 – 1.0 K. This gives the parameters for the fitted data to be $\Gamma_E = 39$ GHz, $\gamma =$ 5.4 MHz, and $T_L = 0.62$ K.